\documentclass{article}
\usepackage{dcase2021,amsmath,graphicx,url,times,booktabs, tabularx, xcolor}
\usepackage{gensymb}
\usepackage{amssymb,amsfonts}

\def \akis [#1]{\textcolor{red}{AP: #1}} 
\def \partha [#1]{\textcolor{blue}{PS: #1}} 

\title{Assessment of Self-Attention on Learned Features For Sound Event Localization and Detection}

\name{Parthasaarathy  Sudarsanam,
      Archontis Politis,
      Konstantinos Drossos
      }
\address{Audio Research Group, Tampere University, Finland\\
        \{parthasaarathy.ariyakulamsudarsanam, archontis.politis, konstantinos.drossos\}@tuni.fi\\
 }

\begin{document}

\ninept
\maketitle

\begin{sloppy}

\begin{abstract}

Joint sound event localization and detection (SELD) is an emerging audio signal processing task adding spatial dimensions to acoustic scene analysis and sound event detection. A popular approach to modeling SELD jointly is using convolutional recurrent neural network (CRNN) models, where CNNs learn high-level features from multi-channel audio input and the RNNs learn temporal relationships from these high-level features. However, RNNs have some drawbacks, such as a limited capability to model long temporal dependencies and slow training and inference times due to their sequential processing nature. Recently, a few SELD studies used multi-head self-attention (MHSA), among other innovations in their models. MHSA and the related transformer networks have shown state-of-the-art performance in various domains. While they can model long temporal dependencies, they can also be parallelized efficiently. In this paper, we study in detail the effect of MHSA on the SELD task. Specifically, we examined the effects of replacing the RNN blocks with self-attention layers. We studied the influence of stacking multiple self-attention blocks, using multiple attention heads in each self-attention block, and the effect of position embeddings and layer normalization. Evaluation on the DCASE 2021 SELD (task 3) development data set shows a significant improvement in all employed metrics compared to the baseline CRNN accompanying the task.

\end{abstract}


\section{Introduction}
\label{sec:intro}

Sound event localization and detection (SELD) is a research problem associated with spatiotemporal analysis of acoustic scenes, providing temporal activity information of target sound classes along with their spatial directions or locations while they are active. The problem has seen increased research activity recently \cite{adavanne2018sound, politis2020overview}, which culminated into the introduction of a new SELD task in the \emph{Detection and Classification of Acoustic Scenes and Events} (DCASE) challenge in 2019, currently on its third iteration\footnote{http://dcase.community/challenge2021/}. The task brings together two long-standing problems in acoustical signal processing: sound event detection (SED) aiming at only a temporal description of target sound classes in the scene, and sound source localization (SSL) aiming at detecting localized sound sources without regard to the type of the emitted sound events. Formulating and addressing the joint problem brings new possibilities in machine listening, robot audition, acoustical monitoring, human-machine interaction, and spatially informed deployment of services, among other applications.

The SELD task has been addressed in literature predominantly with deep learning models, with a few exceptions combining deep-learning SED classifiers with model-based localization \cite{perez2020papafil, nguyen2020sequence}. 
The seminal work of \cite{adavanne2018sound} proposed SELDnet, a model performing both SED and SSL tasks jointly, based on a convolutional and recurrent neural network (CRNN) architecture. 
SELDnet used a series of convolutional layers as feature extractors, operating on multichannel spectrograms, followed by layers of gated recurrent unit (GRU) layers modeling longer temporal context.
Such a CRNN architecture had proved successful in the SED task \cite{cakir2017convolutional}, and was extended in \cite{adavanne2018sound} with a localization inference output branch, predicting the frame-wise direction of arrival (DOA) of each detected class, in a regression manner. 
While alternative architectures have been explored (e.g. ResNets \cite{Cao2020}, TrellisNets \cite{Park2019a}, the R3Dnet of \cite{Shimada2021}), the CRNN architecture has remained the most popular through the submissions in DCASE2019 and DCASE2020.
On the other hand, many innovations were network-independent, focusing on improved input features \cite{Cao2019}, separate modeling of SED and SSL tasks and fusion \cite{Cao2019, nguyen2020sequence}, and improved SELD representations and loss functions \cite{Phan2020, Shimada2021}.

 Recently, the Transformer \cite{NIPS2017_3f5ee243} architecture has shown state-of-the-art performance in a variety of tasks ranging from NLP \cite{NIPS2017_3f5ee243}, to image classification \cite{dosovitskiy2020image} and video object tracking \cite{meinhardt2021trackformer}, among others, and has been proposed as a replacement for both CNNs and RNNs, or combined with convolutional layers in a Conformer \cite{gulati2020conformer} architecture. 
 Transformers base their representational power on self-attention (SA) layers that can model longer temporal or spatial dependencies than typical convolutional layers, while, in contrast to RNNs, they can be efficiently parallelized making them significantly faster during inference. 
 Recently transformers have shown strong state-of-the-art performance in SED tasks \cite{miyazaki2020convolution}, while their use in SSL and SELD proposals has remained limited. Regarding source localization, Schymura et al. integrated self-attention into the outputs of the RNN layers in a CRNN model \cite{schymura2021exploiting} showing performance gains over the standard CRNN. 
 In subsequent work \cite{schymura2021pilot}, RNNs are dropped for transformer layers including linear positional encoding, bringing further performance improvements. 
 With regard to SELD, the first work using SA seems to be the DCASE2020 challenge submission of \cite{Phan2020} which follows a SELDnet-like CRNN architecture, augmented with SA layers following the bidirectional RNN layers. 
 The best performing team in DCASE2020 also seems to employ attention in the form of conformer blocks, as detailed in a later report \cite{wang2021four}. Following DCASE2020, Cao et al. \cite{cao2021improved} proposed their Event Independent Network V2 (EINV2), realizing a track-based output format instead of the class-based one of standard SELDnet, using multi-head self-attention (MHSA) layers following convolutional feature extractors.
 Sinusoidal positional encoding is used before the MHSA as in \cite{NIPS2017_3f5ee243}. 
 Since the above SELD proposals include various other improvements and modifications over the basic SELDnet CRNN, such as modified loss functions \cite{Phan2020}, partially independent models for SED and SSL with parameter sharing \cite{cao2021improved}, or various data augmentation strategies \cite{wang2021four}, the effect of adding self-attention in isolation to the result is not clear.

In this work we exclusively investigate the effects of self-attention in a SELD setting. The rest of this paper is organized as follows. Section 2 presents our baseline method and the multi-head self-attention mechanism. In section 3, we describe in detail our experimental set up used to analyze the effect of self-attention. In section 4, we discuss the results of all our experiments. Finally, in section 5, we present our conclusion of this study.

\section{Method}
\label{sec:method}
For our study, we employ a widely used SELD method that is based on a learnable feature extraction and a learnable temporal pattern identification, that operate in a serial fashion. We call this commonly used SELD method as our baseline. We replace the temporal pattern identification with a self-attention mechanism, that attends to the output of the learnable feature extraction layers.

The input to both the baseline and the version with the self-attention, is a tensor of $K$ sequences of features from different audio channels, each sequence having $T$ feature vectors with $F$ features, $\mathbf{X}\in\mathbb{R}^{K\times T\times F}$. $\mathbf{X}$ is given as an input to the learnable feature extractor. For the baseline, the output of this feature extractor is used as an input to a function that performs temporal pattern identification, and the output of the temporal pattern identification is given as an input to a regressor. In the case of the method used for our study, the output of the learned feature extraction is given as an input to self-attention blocks, and then the output of the latter is given as an input to a regressor. The regressor in both cases predicts the directions-of-arrival for all classes and at each time step, represented by the directions of the output Cartesian vectors. Using the ACCDOA \cite{Shimada2021} representation, the detection activity is also integrated into the same vector representation, with the length of the vectors encoding the probability of each class being active. The output of the regressor and the targets are $\hat{\mathbf{Y}}\in \mathbb{R}^{T\times C\times 3}$ and $\mathbf{Y}\in \mathbb{R}^{T\times C\times 3}$ respectively, where C is the number of classes and 3 represents the Cartesian localization co-ordinates.

\subsection{Baseline}
As the baseline, we use the CRNN architecture proposed in~\cite{Adavanne2018_JSTSP}, with ACCDOA representation for the output. The baseline has three convolutional neural network (CNN) blocks, $\text{CNNBlock}_{n}$ with $n=1, 2, 3$. $\text{CNNBlock}_{n}$  acts as the learnable feature extractor, extracting high level representations from $\mathbf{X}$ as, 
\begin{equation}
    \mathbf{H}_{n} = \text{CNNBlock}_{n}(\mathbf{H}_{n-1})
\end{equation}
\noindent
where $\mathbf{H}_{n}$ is the output of the $n$-th CNN block and $\mathbf{H}_{0} = \mathbf{X}$. Each CNN block consists of a 2D convolution layer, a batch normalization process (BN), a rectified linear unit (ReLU), and a max pooling operation, and process its input as 
\begin{equation}
    \mathbf{H}_{n} = (\text{MP}_{n} \circ \text{ReLU} \circ \text{BN}_{n} \circ \text{2DCNN}_{n})(\mathbf{H}_{n-1})
\end{equation}
\noindent
where $\circ$ indicates function composition. $\text{BN}_{n}$ and $\text{MP}_{n}$ are the batch normalization and max-pooling processes of the $n$-th CNN block, and $\text{2DCNN}_{n}$ is the 2D convolution layer of the $n$-th CNN block. The output of the last CNN block is $\mathbf{H}_{3}\in\mathbb{R}^{T'\times F'}$,  where $T'$ is the time resolution of the annotations and $F'$ is the feature dimension down sampled from input dimension $F$ in the CNNBlocks.

$\mathbf{H}_{3}$ is used as an input to a series of $m$ recurrent neural networks (RNNs), with $m=1, 2$ as 
\begin{equation}
    \mathbf{H}'_{m} = \text{RNN}_{m}(\mathbf{H}'_{m-1})
\end{equation}
\noindent
where $\mathbf{H}'_{m}\in\mathbb{R}^{T'\times F''}$ is the output of the $m$-th RNN, where $F''$ is the hidden size of the RNN and $\mathbf{H}'_{0}= \mathbf{H}_{3}$
 
The output of the RNN blocks is fed to a fully connected layer. The fully connected layer combines the learnt temporal relationships and it is followed by the regressor layer which predicts the detection and direction of arrival for all the classes for each time step in ACCDOA format. 

\begin{equation}
    \mathbf{y'} = \text{FC1}(\mathbf{H'}_{2})\\
\end{equation}
\begin{equation}
    \mathbf{\hat{Y}} = \text{FC2}(\mathbf{y'})\\
\end{equation}

where $\mathbf{\hat{Y}} \in  \mathbb{R}^{T' \times C \times 3}$ is the predicted ouput from the model. 

\subsection{ACCDOA representation}

The annotations in the dataset for detections are of the form $\mathbf{Y}_{det} \in \mathbb{R}^{T' \times C}$, where $T'$ is the number of time frames and C is the number of classes. For each time frame, the value is 1 for a class which is active, 0 otherwise. For localization, the labels are $\mathbf{Y}_{loc} \in \mathbb{R}^{T' \times C \times 3}$, which gives the 3 Cartesian localization co-ordinates for the classes in each time step that the classes are actrive. 

The ACCDOA output representation simplifies these two labels into a single label $\mathbf{Y}\in \mathbb{R}^{T'\times C\times 3}$. In this representation, the detection probalility score is the magnitude of the predicted localization vector. This value is thresholded to predict the detection activity for each class. Thus the need for two different output branches to predict detection and localization separately becomes unnecessary.    

\begin{table*}[t]
    \caption{Detection and localization results for different  configurations of self-attention block on DCASE 2021 Development set. (* - Size of self-attention head in each layer)}
    \hfill \break
    \centering
    \begin{tabular}{c|c|c|c|c|c|c|c|c}
    \hline
         \emph{N} & \multicolumn{1}{|c}{\emph{M}} & \multicolumn{1}{|c}{\emph{P}} & 
        \multicolumn{1}{|c}{\emph{LN}} &
        \multicolumn{1}{|c}{\# params} &
        \multicolumn{1}{|c}{$ER_{20}$} & \multicolumn{1}{|c}{$F_{20}$} & \multicolumn{1}{|c}{$LE_{CD}$} & \multicolumn{1}{|c}{$LR_{CD}$}\\
         \hline \hline
        \multicolumn{4}{c|}{\textbf{Baseline-CRNN}} & 0.5 M & 0.69 & 33.9 & 24.1 & 43.9 \\
       \hline \hline
         1 & 4 & No & No & 0.3 M & $0.65\pm0.01$ & $38.11\pm1.44$ & $23.17\pm0.85$ & $46.73\pm1.44$ \\
        \hline     
        1 & 8 & No & No & 0.6 M & $0.65\pm0.01$ & $39.12\pm1.48$ & $22.78\pm0.73$ & $46.71\pm1.25$ \\
        \hline
        1 & 12 & No & No & 0.9 M & $0.65\pm0.01$ & $38.96\pm1.06$ & $22.96\pm0.88$ & $46.74\pm1.94$  \\
        \hline \hline
       2 & 8 & No & No & 1.1 M & $0.67\pm0.01$ & $36.95\pm1.16$ & $23.44\pm1.27$ & $44.66\pm1.53$ \\
        \hline
       3 & 8 & No & No & 1.6 M & $0.78\pm0.02$ & $19.57\pm3.63$ & $27.05\pm0.90$ & $22.96\pm4.83$ \\
        \hline
        2 & 8 & No & Yes & 1.1 M & $0.62\pm0.01$ & $44.62\pm1.34$ & $22.03\pm0.66$ & $55.04\pm1.34$ \\
        \hline
        3 & 8 & No & Yes & 1.6 M & $0.62\pm0.01$ & $44.11\pm0.74$ & $22.04\pm0.53$ & $54.61\pm1.07$ \\
        \hline
        2 & 12 & No & Yes & 1.6 M & $0.63\pm0.01$ & $43.95\pm0.69$ & $22.13\pm0.36$ & $54.23\pm0.90$ \\
        \hline
        3 & 12 & No & Yes & 2.4 M & $0.64\pm0.01$ & $43.10\pm0.70$ & $22.38\pm0.54$ & $54.00\pm1.49$ \\
         \hline \hline
        3 (128-256-128)* & 8 & No & Yes & 2.2 M & $0.63\pm0.01$ & $44.65\pm1.88$ & $21.98\pm0.51$ & $55.15\pm1.47$\\
        \hline
        3 (128-64-128)* & 8 & No & Yes & 1.4 M & $0.63\pm0.01$ & $43.64\pm1.23$ & $22.06\pm0.46$ & $54.24\pm1.11$ \\
        \hline \hline
        2 & 8 & Yes & Yes & 1.1 M & $\mathbf{0.61\pm0.01}$ & $\mathbf{45.84\pm1.06}$ & $\mathbf{21.51\pm0.74}$ & $54.99\pm1.87$\\
        \hline
        3 & 8 & Yes & Yes & 1.6 M & $0.62\pm0.01$ & $44.63\pm1.14$ & $21.56\pm0.46$ & $54.46\pm0.94$\\
        \hline
        3 (128-256-128)* & 8 & Yes & Yes & 2.2 M & $0.62\pm0.01$ & $45.14\pm1.03$ & $21.67\pm0.41$ & $\mathbf{55.29\pm1.23}$\\
        \hline
   \end{tabular}
    
   \label{tab:results}
\end{table*}

\subsection{Multi-head Self-Attention in SELD}
\label{subsection:experiments}
The motivation of this study is to quantify the effect of replacing the RNN blocks in the baseline with self-attention blocks to capture the temporal relationships. In our experiments, the convolutional feature extractor is kept exactly the same as in the baseline architecture. The output $\mathbf{H_3}$ from the convolutional feature extractor is passed through a series of $N$ self-attention blocks, with $N=1, 2, .. $ as, 
\begin{equation}
    \mathbf{H}'_{N} = \text{SABlock}_{N}{\{M,P,LN\}}(\mathbf{H}'_{N-1})\\
\end{equation}
where $\mathbf{H}'_{N}\in\mathbb{R}^{T'\times F''}$ is the output of the $N$-th self-attention block, where $F''$ is the attention size and $\mathbf{H}'_{0}= \mathbf{H}_{3}$.

 In particular, we systematically study the effects of number of self-attention blocks (\emph{N}), number of attention heads (\emph{M}) in each self-attention block, positional embeddings (\emph{P})) for each time step and the effect of layer normalization (\emph{LN}) on the detection and localization metrics.
 
 The self-attention layer calculates the scaled dot-product attention \cite{NIPS2017_3f5ee243} of each time step in the input with itself. For any input $\mathbf{H} \in \mathbb{R}^{T\times I} $, where $T$ is the number of time steps and $I$ is the input dimension, its self-attention is calculated as,
\begin{equation}
    \text{SA(\textbf{H})} = \text{softmax}\mathbf{(HW_q W_k^TH^T) HW_v}\\
    \label{eq:sa}
\end{equation} 
Here, $\mathbf{W_q, W_k} \in \mathbb{R}^{I\times K}$ and $\mathbf{W_v} \in \mathbb{R}^{I\times O}$ are learnable query, key and value matrices respectively. $K$ is the key dimension in the attention layer and $O$ is the output dimension.
 
 First, we ran experiments to determine the optimal number of attention heads for the task. A single attention head allows each time step to attend only to one other time step in the input. For SELD task, it is useful to attend to more than one timestep to establish semantic relationships in the input audio scene. A multi-head self-attention (MHSA) layer is described as,
 \begin{equation}
    \text{MHSA}\mathbf{(H)} = \underset{m=1,2,..,M}{Concat} [\text{SA}_{m}(\mathbf{H})]\mathbf{W_p}\\
    \label{eq:mhsa}
\end{equation}
 where M is the number of heads. The output from all the heads are concatenated and $\mathbf{W_p} \in \mathbb{R}^{MO \times O}$, a learnt projection matrix projects it into the desired output dimension.
 
 Next, we studied the effect of stacking multi-head self-attention blocks. It enables the model to learn high level temporal features of different time scales. We also experimented with different ways to stack these MHSA blocks. Specifically, we compared the effect of having layer normalization (LN) and residual connections between successive blocks and not having both. The first multi-head self-attention layer takes as input the features from the CNN. The inputs to the successive layers of MHSA are given by,
\begin{equation}
    \mathbf{H}_N = \text{LN}(\text{MHSA}_{(N-1)}(\mathbf{H}_{N-1}) + \mathbf{H}_{N-1} )\\
     \label{eq:stacked-mhsa}
\end{equation}

 At last, we assessed the effect of having position embeddings in the self-attention block. Position embeddings are helpful in keeping track of the position and order of features that occur in an audio scene. This helps the model to learn temporal dependencies based on order of the sound events. Instead of using a sinusoidal position vector originally  proposed in \cite{NIPS2017_3f5ee243}, since the data is split into chunks and the number of time steps is always fixed in our case, we used a fixed size learnable embedding table. If $P \in \mathbb{R}^{T\times I}$ is the position embedding, then the self-attention of input $H$ with position embedding is calculated as $\text{SA}(H+\emph{P})$ in equation (\ref{eq:sa}).

\section{Evaluation}
\label{sec:evalution}

\subsection{Dataset}

We trained and evaluated our models using the dataset provided for the DCASE 2021 sound event localization and detection challenge \cite{politis2021dataset}. The development set contains 600 one-minute audio recordings with corresponding detections belonging to 12 different classes (alarm, crying baby, crash, barking dog, female scream, female speech, footsteps, knocking on door, male scream, male speech, ringing phone, piano)  and their localization labels. 

The multi-channel audio data is available in two recording formats, 4-channel first-order ambisonics (FOA) format and 4-channel tetrahedral microphone recordings (MIC) format. We used the 4-channel FOA recordings with a sampling rate of 24kHz. The audio recordings also contain realistic spatialization and reverberation effects from multiple multi-channel room impulse responses measured in 13 different rooms. The data is split into 6 folds of 100 recordings each. Folds 1-4 are used for training while 5 and 6 are used for validation and evaluation respectively.

\begin{figure}[t]
  \includegraphics[width=80mm, height=100mm ]{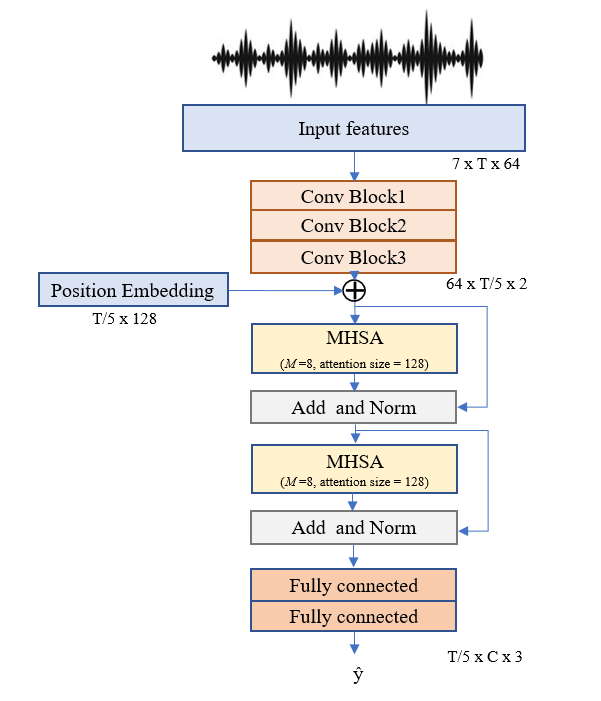}
  \caption{  MHSA model configuration for SELD task.}
  \label{fig:best model config}
\end{figure}

\subsection{Network Training}

As described in section \ref{subsection:experiments}, we analysed the effect of different settings for the self-attention block. First, we replaced the two GRU layers in the baseline, with a single self-attention layer with 4 heads and an attention size of 128. This early result already suggested that using self-attention layers were beneficial compared to RNN layers. With the single layer self-attention, we then set the number of heads to  8 and 12 to evaluate the best hyper-parameter for the number of heads. 

Next, we studied the effect of number of self-attention blocks. Specifically, we modified the architecture to have 2 and 3 attention blocks. For each of these configurations, we also varied the number of heads to be 8 and 12. The self-attention dimension was kept at 128 for all these experiments. When stacking self-attention blocks, we studied the effect of having and not having layer normalization and residual connections between sucessive blocks. In architectures having three self-attention blocks, we also studied the effect of the attention dimension in the multi-head self-attention blocks. In particular, we used 128-128-128, 128-256-128 and 128-64-128 configurations. Finally, we studied the effect of adding positional embedding vectors to the input of the first self-attention layer. We added learnable position embedding of vector size 128 to each time step in the input sequence to the self-attention.  

For all our experiments, as input features, we extracted log mel spectrograms with 64 mel bins for each channel in the multi-channel audio. For the spectrogram extraction, we used short-time Fourier transform (STFT) with a Hann window, 50\% overlap between frames and a hop length of 0.02 seconds. Further, we also calculated the intensity vectors \cite{e536cacb04f141eb9179c6628c23290c} of the multi-channel audio signal from its linear spectra. The log mel spectrograms and the intensity vectors are concatenated along the channel dimension and fed as input to our model. The model is trained for 100 epochs using Adam optimizer with $\beta_1 = 0.9$, $\beta_2 =0.999$ and a learning rate of 0.001. We employed mean squared error as our objective function for this regression task and the model with the best validation score was chosen for evaluation.

The detection metrics are F score and error rate, they are also location-dependent, using a spatial threshold for true positives as detailed in \cite{politis2020overview}. Similar to DCASE2020, true positives occur only if events are localized within 20° from the ground truth of the same class. The localization metrics are localization error and localization recall and they are class dependent. For each setting, we train the model 10 times and report the average scores along with the standard deviation for each metric.

\section{Results}
\label{sec:results}

The results of all our experiments are summarized in Table \ref{tab:results}. Our results from the first set of experiments for determining the appropriate number of attention heads showed that using 8 attention heads was marginally better than 12 heads when the number of attention blocks is fixed to one. Compared to the baseline, the detection error rate decreased from 0.69 to 0.65 and the F score increased from 33.9 to 39.12. There was also a decrease in the localization error from 24.1 to 22.78 and increase in the recall score from 43.9 to 46.71.

Our next set of analysis was to find the optimal number of self-attention blocks.  Experimental results clearly demonstrate that serially connecting more self-attention blocks without layer normalization drastically reduces the performance of the model.  Adding residual connections and layer normalization  between the self-attention blocks significantly improves the performance of the model. We also verified that with multiple self-attention blocks, 8 attention heads was still the best performing configuration. With two self-attention blocks and 8 heads each, there was a steep increase in the F score to 44.62 and the localization recall jumped to 55.04.

Finally, we examined the importance of position embeddings to the first self-attention block and it proved to further increase the performance of our SELD system.  From all our experiments, the best model configuration had two self-attention blocks with eight attention heads each with an attention dimension of 128, a learnt fixed size position embedding and  residual connections with layer normalization between successive self-attention blocks. For this configuration, the detection error rate \textbf{$ER_{20}$} (lower the better), decreased by 11.6\%  and F-score \textbf{$F_{20}$} (higher the better), increased by 35.2\% compared to the baseline. Similarly, the localization error rate \textbf{$LE_{CD}$}(lower the better) reduced by 10.7\% and the localization recall \textbf{$LR_{CD}$} (higher the better) improved by 25.2\% from the baseline. This model configuration is shown in Figure \ref{fig:best model config}.

The best model configuration has close to twice the number of parameters as the baseline. However, due to the parallelization achieved by the self-attention blocks, it is also 2.5x faster than the baseline model during inference, based on our experiments on a V100 GPU. 
Hence, MHSA based models can be useful over RNN based models for real-time SELD tasks.

\section{Conclusions}
\label{sec:conclusions}

In this study, we systematically assessed the effect of self-attention layers for the joint task of sound event detection and localization. To account only for the impact of self-attention on this task, we employed the common SELDnet model using CRNN architecture and studied the effects of replacing the temporal pattern recognition RNN blocks with self-attention blocks. We experimented with various hyper parameter settings for the self-attention block such as number of blocks, number of attention heads in each self-attention block, size of the attention, layer normalization and residual connections between sucessive self-attention blocks and adding positional embedding to the input of self-attention block. Our experiments showed that, multi-head self-attention blocks with layer normalization and position embeddings significantly improve the $F_{20}$ score and $LR_{CD}$ score compared to the baseline. There is also a considerable decrease in the detection and localization error metrics compared to the baseline. The self-attention blocks also  reduced the time required for training and inference compared to RNN blocks by exploiting parallel computations.  

\section{ACKNOWLEDGMENT}
\label{sec:ack}
The authors wish to acknowledge CSC-IT Center for Science, Finland, for computational resources. K. Drossos has received funding from the European Union’s Horizon 2020 research and innovation programme under grant agreement No 957337, project MARVEL.

\bibliographystyle{IEEEtran}
\bibliography{refsAP}

%
%
%
%
%
%
%
%
%

\end{sloppy}
\end{document}